# Real-Time Imaging of Quantum Entanglement

Robert Fickler[1,2,*], Mario Krenn[1,2], Radek Lapkiewicz[1,2], Sven Ramelow[1,2], Anton Zeilinger[1,2,3,*]

[1]Quantum Optics, Quantum Nanophysics, Quantum Information, University of Vienna, Boltzmanngasse 5, Vienna A-1090, Austria

[2]Institute for Quantum Optics and Quantum Information, Austrian Academy of Science, Boltzmanngasse 3, Vienna A-1090, Austria

[3]Vienna Center for Quantum Science and Technology, Faculty of Physics, University of Vienna, Boltzmanngasse 5, Vienna A-1090, Austria

*Correspondence to: robert.fickler@univie.ac.at and anton.zeilinger@univie.ac.at

Quantum Entanglement[1,2] is widely regarded as one of the most prominent features of quantum mechanics and quantum information science[3]. Although, photonic entanglement is routinely studied in many experiments nowadays, its signature has been out of the grasp for real-time imaging. Here we show that modern technology, namely triggered intensified charge coupled device (ICCD) cameras are fast and sensitive enough to image in real-time the effect of the measurement of one photon on its entangled partner. To quantitatively verify the non-classicality of the measurements we determine the detected photon number and error margin from the registered intensity image within a certain region. Additionally, the use of the ICCD camera allows us to demonstrate the high flexibility of the setup in creating any desired spatial-mode entanglement, which suggests as well that visual imaging in quantum optics not only provides a better intuitive understanding of entanglement but will improve applications of quantum science.

A fairly young but vibrant field that studies the spatial structure of the optical modes of photons (e.g. Laguerre-Gauss[4], Ince-Gauss[5], Bessel-Gauss[6]) continues to attract wide interest. Each spatial mode offers many interesting features, like orbital angular momentum[7] or continuous vortex splitting[8], which already lead to novel insights in quantum optics like higher dimensional entanglement[9,10,11,12], novel uncertainty relations for the angular and OAM degree-of-freedom[13,14], remote object identification[15] or angular sensitivity enhancement with very high OAM[16].

The rapid progress in imaging technologies over the last few years has made CCD cameras an interesting option for single photon detection in quantum optics experiments, since the spatial information is directly accessible. Due to high detection efficiencies, electron multiplied CCD cameras have attracted attention recently and have been used to show non-classical correlations from photons produced via spontaneous parametric down conversion (SPDC)[17,18,19,20]. The downside of such cameras is that they only allow relatively long exposure times (μs) which makes it necessary to sum over many images with a sparse number of photons and makes it unfeasible to use them for coincidence imaging of entanglement. In contrast, ICCD cameras have lower quantum efficiencies due to the intensifier and fluorescence screen in front of the CCD chip but show a very good signal to noise ratio and therefore good single photon sensitivity. They have been used to illustrate non-classical effects of the photons from the SPDC process [21,22,23,24,25]. However, the biggest advantage of ICCD cameras is the very fast (~2ns) and precise (~10ps) optical gating of the intensifier which makes it possible to use them in a coincidence scheme for real-time imaging of quantum entanglement.

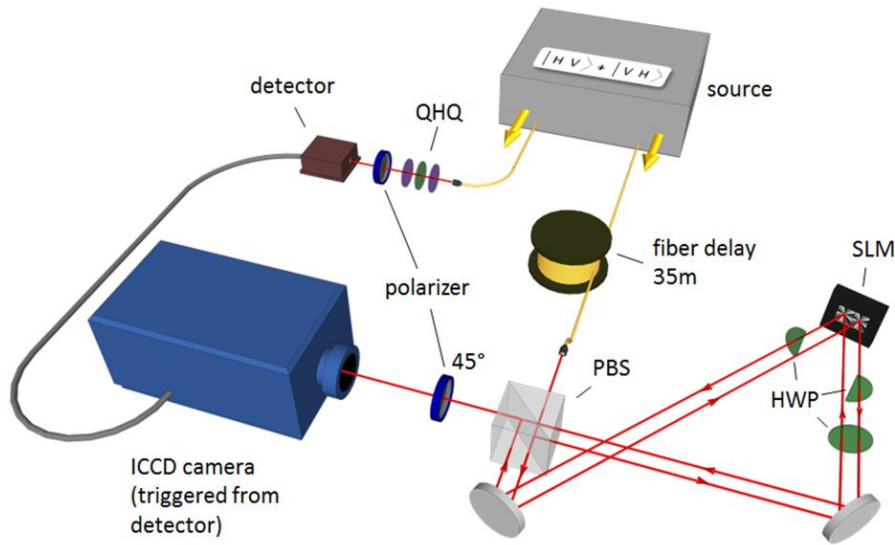

Figure 1: Sketch of the experimental setup. Polarization entanglement is created in a SPDC process (source – grey box) and both photons are coupled into single mode fibers (yellow). One photon is measured in the polarization bases with a combination of two quarter wave plates, a half wave plate (QHQ – violet and green) and a polarizer (blue). The photon is detected by single photon detector (brown) and the detector signal is used as a trigger for the ICCD camera. The second photon is delayed by a 35m fiber (to account for the delay time from the trigger detector, the travel time of the trigger signal and insertion delay from the ICCD) and brought into the interferometric transfer setup, which consists of a polarizing beam splitter (PBS), three half wave plates (HWP – green), a spatial light modulator (SLM - black) and a polarizer (blue) at 45° after the interferometer. In the transfer setup the HWPs rotate polarization to ensure the optimal working of the SLM and to separate the output from the input path. Depending on the polarization the photon gets transferred by the SLM to any desired spatial mode. The polarizer after the interferometer erases any information to which spatial mode the photon was modulated and thus completes the transformation. The spatial mode of the photon is registered by the triggered ICCD camera which is gated for 5ns and therefore only detects the transferred photons which belong to polarization encoded trigger photons.

In our experiment, we use a combination of the polarization and spatial degree-of-freedom (DOF) to be able to directly image entanglement. We start with a high-fidelity polarization-entangled two-photon state (Figure 1). One photon is unchanged; the other photon is brought to a second setup, which transfers the polarization DOF to a wide range of specifically chosen spatial mode. In this interferometric setup[16], the photons get transferred by a liquid-crystal spatial light modulator, dependent on their polarization (methods), to a hybrid-entangled two-photon state

$$|\psi\rangle = \alpha|H\rangle|spM_1\rangle + e^{i\phi}\beta|V\rangle|spM_2\rangle, \qquad (1)$$

where $\alpha$, $\beta$, and $\Phi$ are real and $\alpha^2 + \beta^2 = 1$, $H$ and $V$ denote the horizontal and vertical polarization, $spM_1$ and $spM_2$ correspond to arbitrary spatial modes, and the positions of the ket-vectors label the different photons. In order to image the created spatial mode and demonstrate entanglement between the two photons, the polarization encoded photon is projected onto a certain polarization and detected by a single photon detector. The signal from the detector is used as a trigger for the ICCD camera, which in turn registers the transferred photon.

In our measurements, we use an ICCD (Andor iStar A-DH334T-18F-03) with a quantum efficiency of 3% for 810nm wavelength, a gating of 5ns and a spatial resolution of 1024x1024 pixels (effective pixel size: 13x13µm). With this camera we observe clear single-photon images even for very complex mode structures (Figure 2 c and Figure 4), where the whole spatial information is directly available with a very high precision. Compared to scanning or masking of single-pixel detectors, direct imaging with an ICCD also shortens the measurement time significantly. Thus, the advantage of real-time imaging with an ICCD is an improvement - both spatially and temporally - of many orders of magnitude, opening up possible novel applications in quantum information and quantum metrology. Note that similar ICCD cameras with 20% efficiency and 2ns gating, which are readily available, promise a 20-fold increase in the signal to noise ratio. Since the adjustable insertion delay time for triggering the ICCD is at least 35ns, we delay the second photon with a fibre before sending it through the transfer setup. If a wrong delay is chosen nearly no accidental photon events can be seen at the camera (see Figure 2 b) and hence no background correction has to be applied. This suggests the possibility for precise measurements in the temporal domain. A few residual events appear due to the high triggering rate (MHz) and the resulting thermal noise from the intensifier and from the afterglowing of the fluorescing phosphorous screen. We find the ratio between the number of detected photon events in a picture (right delay) to the residual events (wrong delay) to be on the order of 75:1. Those undesirable residual events may be suppressed substantially, with more efficient ICCD cameras and/or smaller gating times.

To image the effect of entanglement, we scan the Poincaré sphere of the polarization encoded trigger photon and register the appearing mode pattern from the transferred photons at the ICCD camera. In this way, we are able to visualize directly the probability distribution of complex spatial modes of the whole Bloch sphere (Figure 2 a). Since the measurement of each spatial mode only takes a few seconds (e.g. 3 seconds for $LG_{\pm 1}$), the influence of the polarization measurement of the first photon is visible in real-time at the ICCD camera (Supplementary Movie 1). Entanglement is already visible in this video, since the high-contrast minima and maxima shift in very good correspondence to the polarization angle measured on the partner photon.

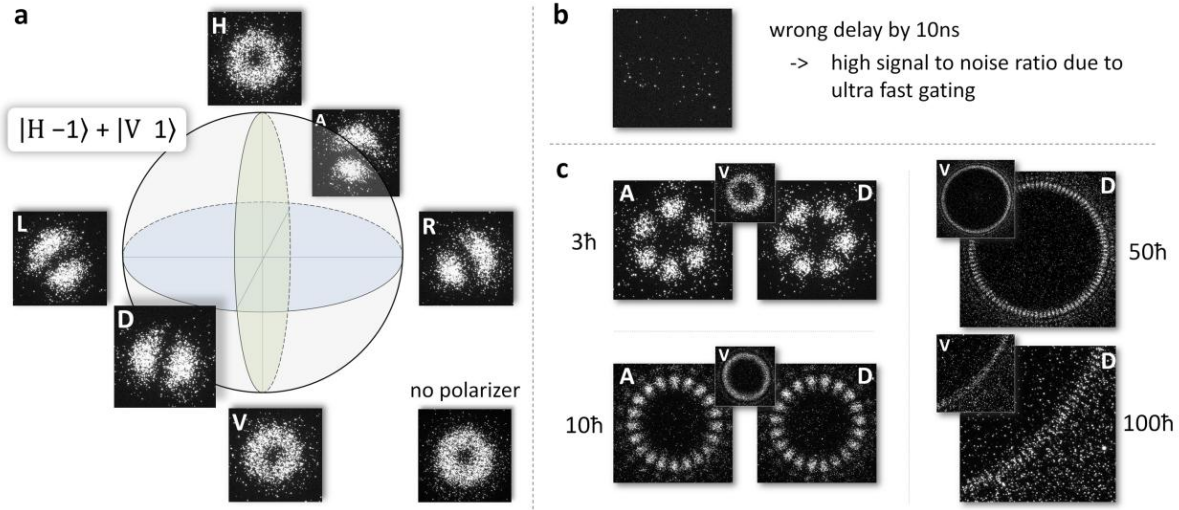

Figure 2: Gallery of single photon images where the photons are transferred to different orders of LG modes. Depending on which polarization the entangled trigger photon is projected (white letters in the images), different mode patterns are registered with ICCD camera. (a) shows a Bloch sphere for the first order Laguerre-Gauss modes. By scanning the polarization encoded trigger photon around its Poincaré sphere, the whole LG-Bloch sphere of the entangled partner photon in the spatial mode can be visualized in real-time. A sequence of single-photon images for different trigger polarizations around a meridian (green circle in the sphere) and the equator (blue circle in the sphere) confirms directly the presence of entanglement due to the high-contrast minima and maxima for the two mutually unbiased bases and can be seen in the Supplementary movie 1. If no polarizer is put in the path of the trigger photon (bottom right) a statistical mixture of all states of the LG-Bloch sphere is registered. (b) If the delay is changed to a wrong value by 10ns, the gating time of the ICCD camera does not match the arrival of the delayed photons and nearly no intensity is registered. This demonstrates the high signal to noise ratio and the capability of high temporal resolution. (c) Although the structure of the superposition for higher order LG modes becomes more complicated and the resolution of the SLM and the camera is getting crucial, the characteristic petal structure can be identified even up to the $100^{th}$ order. Note that at the camera one photon event is distributed over more pixels of the CCD because each channel of the intensifier is bigger than the CCD pixels and therefore is spread over many pixels.

While visual observation already intuitively confirms the presence of entanglement, we also verify it quantitatively: Since the registered signal of the camera depends linearly on the detected photon number, we determine the average signal per detected photon and its error margin from many single photon events (Supplementary Information). With this relation between registered signal and corresponding photon number it is possible to spatially analyse any recorded intensity image without the need for individual counting of single photons over a time consuming data acquisition of many sparse images.

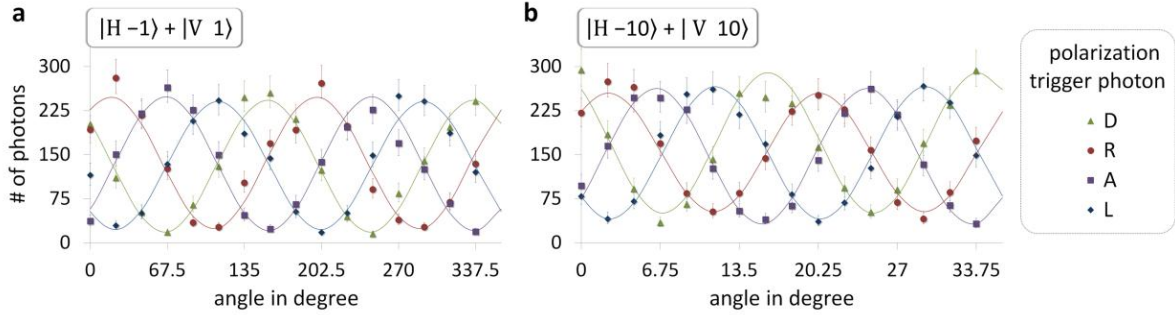

Figure 3: Detected photon number per angular region for different polarizations of the trigger photon. From the recorded images (Figure 2) we evaluate the number of photons per angular region (bin size $\frac{22.5°}{l}$) and thereby verify quantitatively entanglement. In (a) and (b) non-classical fringes for hybrid entanglement between polarization and LG modes with $l=1$ and $l=10$ are shown respectively. Error bars are obtained from a Monte Carlo simulation which is based on the statistics of around 5800 single photon events (Supplementary Information). From the measured minima and maxima of the photon numbers, the visibilities are calculated and used to violate the separability bound of an entanglement witness. Because of the periodic structure of LG superpositions all angular regions in multiples of $\frac{360°}{l}$ are summed up to get a bigger number of photons per angle and therefore a better statistical significance.

To confirm entanglement we make use of a specific feature of the Laguerre-Gauss (LG) mode family. The spatial structure of two superimposed LG modes with opposite helicities $|LG_{\pm l}\rangle = |LG_l\rangle + e^{i\varphi}|LG_{-l}\rangle$ shows a radially symmetric distribution where a change of the phase $\varphi$ between the two modes directly translates to a spatial rotation of $\frac{\varphi}{2l}\frac{360°}{2\pi}$. To discriminate between different orientations of the structure and therefore different superpositions, we evaluate the photon number per angular region from the measured intensity image for different trigger polarizations (Figure 3 a and b). From the maximal and minimal detected photon numbers the visibilities in two mutually unbiased bases and therefore the expectation value of an entanglement witness operator $\widehat{W}$ can be calculated[26] (Supplementary Information). For all separable states the inequality

$$\widehat{W} = vis_{D/A} + vis_{R/L} \leq 1 \qquad (2)$$

holds and surpassing this bound verifies entanglement. Capital letters stand for the polarization of the trigger photon (D = diagonal, A = anti-diagonal, R = right circular, L = left circular). For first order LG modes with *l*=±1 we obtained a value of 1.68±0.03 which violates the inequality (2) by more than 20 standard deviations, therefore proving entanglement. For $LG_{\pm2}$, $LG_{\pm3}$, $LG_{\pm5}$ and $LG_{\pm10}$ the measured witnesses are 1.53±0.05, 1.50± 0.05, 1.50± 0.04, and 1.46±0.05 respectively and thus violate the bound for separable states by around 10 standard deviations. We note that no background subtraction was applied, but the measured photon numbers from the registered signal of the ICCD might be a bit smaller than they were in the actual measurement, due to saturation effects where a lot of photons are registered in the same region of the camera, namely the maxima. However, a bigger actual photon number in the maximum would correspond to a higher value of the visibility and therefore a stronger violation than the one presented here.

Recently, it was shown that hybrid-entangled two-photon states of higher-order LG modes can be used to improve sensitivity in the remote sensing of an angular rotation[16]. By using an ICCD camera to image the mode patterns it is possible to visualize this gear-like behaviour between the rotation of the polarization and the petal structure of the spatial mode without any masking and its inherent significant reduction in count rates. In contrast to the experiment in Ref. 16, this significantly shortens the acquisition time. A scan of the polarization around the equator of the trigger photons' Poincare sphere leads to a rotation of the structure by 180° for $LG_{\pm1}$, 90° for $LG_{\pm2}$ and 36° for $LG_{\pm5}$ (Supplementary Movie 2).

Furthermore, the capability of the ICCD camera of resolving complex spatial pattern with a very high precision enables the demonstration of the high flexibility of the presented setup. If transferring one of the photons to the Hermite-Gauss (HG) mode family or the general family of Ince-Gauss (IG) modes, all registered single-photon images show a very good agreement with the theoretical prediction (Figure 4 a and b). Additionally, it is possible to create entanglement between polarization and an artificial mixture of two different mode families at the same time, here a superposition of a higher order LG and higher order HG mode (Figure 4 c). Since no mask is required the imaging with an ICCD camera is a very general way to measure entanglement of any spatial mode or complex pattern of single photons which might advance quantum optics experiments where information is encoded in the spatial domain.

Our results represent the first imaging of entanglement in real-time, where the influence of the measurement of one system on its entangled, distant partner system is directly visible. The use of an ICCD camera to evaluate the number of photons from a registered intensity within a given region opens up new experimental possibilities to determine more efficiently the structure and properties of spatial modes from only single intensity images. The presented results suggest that triggered ICCD cameras will advance quantum optics and quantum information experiments where complex structures of single photons need to be investigated with high spatio-temporal resolution.

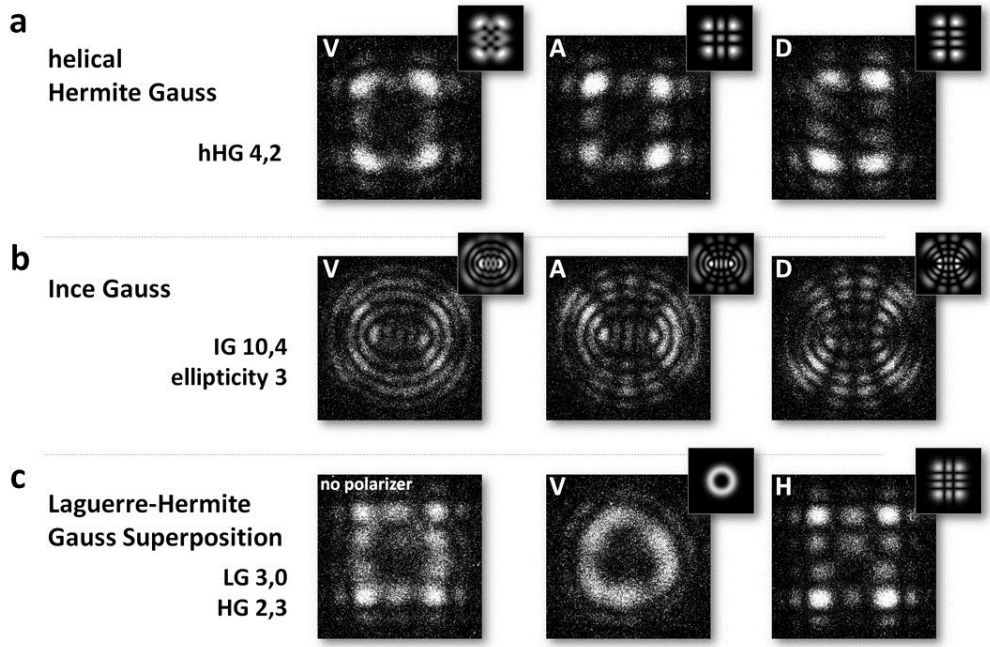

Figure 4: Gallery of registered single photon images where the transferred photon is encoded in different HG, IG, and LG/HG modes which demonstrates the flexibility of the transfer setup in creating any desired spatial mode entanglement. The white letters in each image denote the polarization the entangled trigger photon was projected onto and the small insets above each image correspond to the theoretical expected intensity structure of the transferred photon. (a) For the helical HG mode family[29] a trigger photon in diagonal (D) or anti-diagonal (A) polarization leads to different orders of HG modes for the images of the entangled partner photon. (b) If the photon is transferred to the general family of IG modes characteristic properties like splitting of the vortices (vertical (V) polarized trigger photons) or additional nodal lines (D or A polarized trigger photons) can be seen. (c) The artificial superposition between the mode families of LG with OAM and HG without OAM shows that any custom tailored spatial mode entanglement can be realized.

**Methods**

Source and spatial light modulator (SLM):
The polarization-entangled photon pairs were created in a SPDC process using a 15mm-long type-II nonlinear crystal (periodically poled potassium titanyl phosphate (ppKTP)) in a Sagnac-type configuration[27,28]. A blue 405nm continuous-wave diode laser with up to 35mW of power pumps the crystal and thereby creates photon pairs of 810nm wavelength. Two 3nm band-pass filters were used before the photons were coupled into single-mode fibers. With this approximately 1.3 million pairs per second can be detected at full pump power. Any polarization change between the source and the transfer setup is undone by fiber polarization controllers. The SLM (resolution: 1920x1080, pixel size: 8μm, Holoeye Photonics AG) in the transfer setup, which modulates only the phase of the light, was used to create the desired spatial modes.

**Acknowledgements**

The authors thank T. Pieper and LOT-QuantumDesign for providing the camera. This work was supported by the European Research Council (advanced grant QIT4QAD, 227844) and the Austrian Science Fund (FWF) through the Special Research Program (SFB) Foundations and Applications of Quantum Science (FoQuS; Project No. F4006-N16) and W1210-2 (Vienna Doctoral Program on Complex Quantum Systems; CoQuS).

**Supplementary Material**

Movie 1 and movie 2 files attached

Evaluation of the photon number

The registered intensity at the ICCD camera is read out as a signal in counts per pixel, which does linearly depend on a specific photon number. In order to evaluate the photon number corresponding to a registered intensity, we need to know the average signal caused by a single photon. For this purpose we analyze around 5800 detected single photon events to get a statistically significant mean value. In each shot we subtracted at first the camera-induced readout noise (mean background) of each pixel. In a second step, we summed up all signal counts of a contiguous pixel array as one photon where at least one pixel value is more than 5 standard deviations above the background fluctuations. This has to be done since the photons are spread over a few pixels due to different resolutions of the intensifier and the CCD pixel size (see insets in Supplementary Figure 1a). With the mean value it is now possible to determine the number of photons which correspond to detected signal counts within a certain region of the intensity image. To evaluate the error margin for each photon number, we performed a Monte Carlo simulation based on the obtained probability distribution from the single photon measurements (Supplementary Figure 1a). A very good fit to the resulting histogram of the distribution was found to be a log-normal probability function. With this distribution 50000 possible signal counts were simulated for each photon number. The resulting average signal for every photon number corresponds to the one obtained with the mean value from the single photon events. The thereby found standard deviation can now be used as a look-up table to determine the error margin of each photon number. The linear dependence of registered signal on the photon number as well as the standard deviation is shown in Supplementary Figure 1b and was used to demonstrate the non-classical behaviour (Figure 3 a and b in the main text) and demonstrate entanglement quantitatively by violating the bound of an entanglement witness (equation (2)).

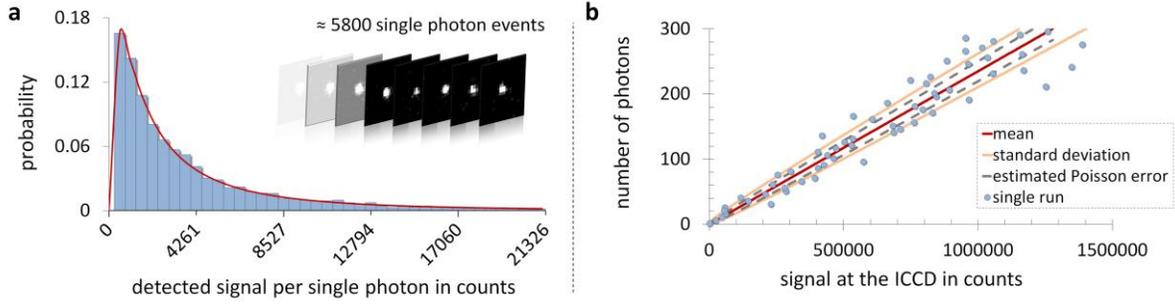

Supplementary Figure 1: (a) shows the measured histogram of the detected signal at the ICCD for around 5800 single photon events. The insets show example images of single photon events. The spread of a single photon over a few pixels is caused by the different resolutions of the intensifier and the CCD chip of the ICCD camera. The log-normal distribution was fitted to the data which is specific to the ICCD camera and was used to Monte Carlo simulate the error margin of each photon number. The linear relation between the signal at the ICCD and the photon number can be seen in (b). With the help of the fitted log-normal probability function 50000 Monte Carlo simulations (runs) were performed for each photon number and the mean value (red line) and the standard deviation (orange line) were obtained. The standard deviation was used to determine the error margin for each photon number. For a comparison, the error estimation from Poissionian statistics is shown (grey dashed line) as well as the result from a single simulation run (blue circles) for every $5^{th}$ photon number.

Calculation of the entanglement witness

The introduced entanglement witness consists of the sum of two visibilities *vis*

$$\widehat{W} = vis_{D/A} + vis_{R/L} \,, \tag{3}$$

where the indices describe the polarization of the trigger photon. It can be calculated as follows: Because the phase between the LG mode with a positive OAM quantum number $l$ and negative OAM value $-l$ is directly connected to the angular position via the formula $\gamma = \frac{\varphi}{2l}\frac{360°}{2\pi}$, it is possible to distinguish between any equally weighted superposition of Laguerre-Gauss states

$$|LG_{\pm l}\rangle = \frac{1}{\sqrt{2}}|LG_l\rangle + e^{i\varphi}|LG_{-l}\rangle \tag{4}$$

with the ICCD camera. The two visibilities of the witness operator can be rewritten in terms of four projections onto different trigger polarizations and angular positions. Hence, the witness becomes

$$\widehat{W} = \left(\widehat{P}_{D,\gamma 1} + \widehat{P}_{A,\gamma 1^\perp} - \widehat{P}_{D,\gamma 1^\perp} - \widehat{P}_{A,\gamma 1}\right) + \left(\widehat{P}_{R,\gamma 2} + \widehat{P}_{L,\gamma 2^\perp} - \widehat{P}_{R,\gamma 2^\perp} - \widehat{P}_{L,\gamma 2}\right) \,, \tag{5}$$

where the position of the indices label the two photons, the capital letters stand for the trigger polarization and the $\gamma$ stands for the angular position of the registered structure at ICCD. $\gamma 1$ can be chosen to fit to the maximum intensity for D polarized trigger photons and thus fixes all following angular positions of the LG superposition measurements. The angle $\gamma 2 = \gamma 1 + \frac{45°}{l}$ therefore corresponds to the second mutually unbiased basis which is needed to verify entanglement. The ⊥-sign illustrates the angular position of the respective orthogonal superposition e.g. $\gamma 1^\perp = \gamma 1 + \frac{90°}{l}$ which is necessary to measure the visibility. To find the bound for all separable states we use the general pure separable two photon OAM state where the first photon is in the polarization mode and the second photon in the LG mode

$$|\psi\rangle = \left(a|H\rangle + b \cdot e^{i\varphi_1}|V\rangle\right) \otimes \left(c|LG_l\rangle + d \cdot e^{i\varphi_2}|LG_{-l}\rangle\right), \tag{6}$$

with $a, b, c, d, \varphi_1, \varphi_2 \in \mathbb{R}$, $a^2 + b^2 = 1$, $c^2 + d^2 = 1$ and $l$ denotes the quanta of OAM. The straightforward calculation of the witness (5) for the separable state (6) leads to

$$\widehat{W} = 4 \cdot a \cdot b \cdot c \cdot d \cdot \cos(\varphi_1 - \varphi_2). \tag{7}$$

Therefore, the maximal value of the witness $\widehat{W}$ is 1 for $a=b=c=d=\frac{1}{\sqrt{2}}$ and $\varphi_1 = \varphi_2$. If the sum of the visibilities is bigger than 1 the measured state is non-separable or in other words entangled:

$$\widehat{W} = vis_{D/A} + vis_{R/L} \begin{cases} \leq 1 & separable \\ > 1 & entangled \end{cases}$$

Note that the presented witness is linear and therefore the calculated bound holds for all separable mixed states as well.

12